# Statistical tests for comparing the associations of multiple exposures with a common outcome in Cox proportional hazard models

Rikuta Hamaya, MD, MSc[1,2], Peilu Wang, MD, PhD[3], Lin Ge, PhD[1,4], Edward L. Giovannucci, MD, ScD[1,5], Molin Wang, PhD[1,4,6]

[1] Department of Epidemiology, Harvard T.H. Chan School of Public Health, Boston, MA
[2] Division of Preventive Medicine, Department of Medicine, Brigham and Women's Hospital and Harvard Medical School, Boston, MA
[3] Department of Nutrition and Food Hygiene, School of Public Health, Institute of Nutrition, Fudan University, Shanghai, China
[4] Department of Biostatistics, Harvard T.H. Chan School of Public Health, Boston, MA
[5] Department of Nutrition, Harvard T.H. Chan School of Public Health, Boston, MA
[6] Channing Division of Network Medicine, Department of Medicine, Brigham and Women's Hospital and Harvard Medical School, Boston, MA

**Address for correspondence:**
Molin Wang, PhD
Departments of Epidemiology and Biostatistics, Harvard T.H. Chan School of Public Health
667 Huntington Ave, Boston, MA, USA
Tel: 617 432 1843
Fax: 617 432 2435
Email: stmow@channing.harvard.edu

**Funding:** This research was supported by grants from National cancer institute (NCI; U01 167552). EG is funded as an American Cancer Society Clinical Research Professors (grant CRP-23-1014041).

**Conflict of interests:** no support from any organization for the submitted work; no financial relationships with any organizations that might have an interest in the submitted work in the

**Abstract**

With advancement of medicine, alternative exposures or interventions are emerging with respect to a common outcome, and there are needs to formally test the difference in the associations of multiple exposures. We propose a duplication method-based multivariate Wald test in the Cox proportional hazard regression analyses to test the difference in the associations of multiple exposures with a same outcome. The proposed method applies to linear or categorical exposures. To illustrate our method, we applied our method to compare the associations between alignment to two different dietary patterns, either as continuous or quartile exposures, and incident chronic diseases, defined as a composite of CVD, cancer, and diabetes, in the Health Professional Follow-up Study. Relevant sample codes in R that implement the proposed approach are provided. The proposed duplication-method-based approach offers a flexible, formal statistical test of multiple exposures for the common outcome with minimal assumptions.

**Introduction**

Epidemiological investigations frequently employ regression models to determine the relationship between a specific exposure and an outcome. With advancement in medicine and data science, alternative forms of exposure, potentially exerting different associations on the identical outcomes, are emerging. The alternative may be the one that measures the same exposure in a different method or a conceptually different exposure but the association with a common outcome is of interest. Consequently, there is a need to discern a difference in the association between an exposure and the alternatives with respect to an outcome. There are multiple studies that compare the associations of alternative exposures with the same outcome. The examples include evaluating three measures of forest fire smoke exposure with respiratory and cardiovascular health outcomes[1], assessing alternative Ozone measures with agricultural crops[2], addressing alternative metrics for ambient traffic-related and regional pollutants for use with ambient air pollution and acute morbidity via ED visits.[3] Most of these studies employed separate regression-based analyses to examine the association between the specific exposure and the target outcome. They also adopted the commonly used univariate Wald test to compare the exposure-outcome associations among different alternative exposures where the associations are estimated in separate analyses based on the same dataset. One potential limitation of this approach is that it does not take account of the correlations between those estimated associations from the separate analysis. Substitution analysis is another popular method to compare alternative exposures especially in nutritional epidemiology, in which two or more exposures are mutually adjusted in the same model.[4] Although the approach also compares two effect estimates, it focuses on the scientific questions regarding whether replacing one exposure with a second exposure would potentially change the outcome. Of interest in this paper is different – whether the exposure-outcome associations are different across the multiple exposures when the exposures are not mutually adjusted.

This paper proposes a data duplication strategy for formally testing whether the associations of multiple exposures with an outcome are differential. In our motivating study detailed in a later section, we focused on two representative dietary patterns which were derived based on predefined dietary components (the Alternative Healthy Eating Index— 2010 [AHEI-2010][5]) or based on empirical analyses (reversed empirical dietary inflammatory pattern [rEDIP][6]) with respect to the association with incidence of chronic diseases. Given that there are correlations between the estimated coefficients across multiple exposures, it is crucial to employ an appropriate statistical test to assess differences in the associations between these exposures regarding the same outcome, while accounting for the existing correlations. The proposed approach provides valuable insights for clinical practice and epidemiological evidence, and it can also be applicable for other regression-based models.

**Models and Methods**

We consider a setting of cohort study in which the objective is to compare the association of exposure $A$ versus alternative exposure $A'$ with the same outcome. Here we assume two exposures for presentational simplicity, while the extension of this approach to more than two exposures is straightforward. The exposures can be modelled as continuous, dichotomous, or categorical variable. In case that $A$ is categorical variable, we assume that the categories of $A'$ is accordant with those of $A$; e.g. both $A'$ and $A$ are based on quintiles. In addition, we assume that a list of covariates $L$ is the same for $A$ and $A'$ with respect to the same outcome $Y$. If distinct covariate lists exist, the more comprehensive list including covariates from both models should be used in the two models. For the statistical analysis in this article, we assume the following Cox proportional hazard models:

$$\lambda_1(t|A, L) = \lambda_{01}(t) \exp(\boldsymbol{\beta_1}^T A + \boldsymbol{\alpha_1}^T L)$$

$$\lambda_2(t|A', L) = \lambda_{02}(t) \exp(\boldsymbol{\beta_2}^T A' + \boldsymbol{\alpha_2}^T L),$$

where $\lambda_i(t)$ and $\lambda_{0i}(t)$ denote hazard and baseline hazard functions at time t, respectively; $\boldsymbol{\beta_i}$ and $\boldsymbol{\alpha_i}$ are the coefficients for exposure(s) and covariate(s), respectively and $i = 1, 2$ denotes two separate analyses. Our aim is to test the following null hypothesis:

$$H_0: \boldsymbol{\beta_1} = \boldsymbol{\beta_2}.$$

For a cohort study, age is often used as the time scale as it is typically the most important confounder in chronic disease epidemiology[7], and the counting process data structure can be used to deal with time-dependent covariates and left truncation. The proposed method can be readily adapted to alternative study design.

### *Dichotomous or single continuous exposures*

The proposed method is inspired from the duplication method[8] and implemented similarly to the statistical methods for studying disease subtype heterogeneity[9]. First, we aim to perform valid statistical test for the difference of two dichotomous or continuous variables with respect to one outcome in a Cox proportional hazard model. Our proposed strategy is to firstly duplicate the dataset in a row binding manner and then create two new variables: Exposures, which contains the values of $A$ and $A'$ in the original dataset; A_type, which is an indicator for the exposures ($A$ or $A'$). The original variables of $A$ and $A'$ will not be used in the new augmented dataset. See **Figure 1** for an example of the augmented data creation.

A Cox proportional hazard model is then fitted on the augmented data, stratified by A_type and including Exposures, all covariates, and the interactions between A_type and Exposures as well as between A_type and each covariate. Robust standard errors are specified in the model to account for potential correlations due to the duplication of same individuals labelled with `id`. Theoretically, when fitting the Cox model for the augmented dataset, the maximum partial likelihood method-based point estimates of exposure coefficients are the same as those obtained in separate analysis.[9] The only difference is the estimation for the variance-covariance matrix of the exposure coefficients, where the proposed approach will include the estimated covariances between the coefficients of alternative exposures with respect to the same outcome.

Finally, a Wald test examining whether the coefficient for the interaction between A_type and Exposures is zero serves as the statistical test for the difference in the associational patterns between multiple exposures for the same outcome. When we are applying this method for single continuous exposure, the test is essentially for the difference in the linear associations with the log-hazard ratios for the incidence of the outcome event.

### *Exposures with ≥3 categories or multiple continuous exposures*

In case when examining two exposures with ≥3 categories or when testing more than two continuous exposures is interested, a multivariate Wald test may be employed for this purpose. For exposures, with $n \geq 3$ categories, a total of $(n-1)$ dummy variables may need to be created, although this is not necessary when using factor variables in R. Similar to the scenario with dichotomous exposures, the original dataset is duplicated for a new augmented one, and the indicator A_type variable is created to differentiate two analyses (see the example in **Figure 2**); in this instance, Exposures variables will be generated for each dummy variable (Exposures2 and Exposures3 in **Figure 2**). Subsequently, the new augmented dataset is utilized to fit a Cox proportional hazard model, which is stratified by A_type and includes all Exposures, all covariates, the interactions between A_type and each covariate, and the interactions between A_type and each exposure. Finally, a multivariate Wald test is conducted for the Cox proportional hazard model. The null hypothesis being tested posits that the coefficients of the interactions between Exposures and A_type are all zeros. The degree of freedom in the multivariate Wald test corresponds to the number of comparisons in the statistical hypothesis, i.e., the number of the interactions between A_type and exposures. To compare the linear trends in associations over ordinal categories (e.g., quintiles) among two or more exposures, for each exposure, the median value in each ordinal category can be treated as a continuous variable. The test is then the same as for comparing continuous exposures.

**Examples with R codes**

We illustrate the aforementioned methods through a real-world epidemiological study. Several dietary patterns have been recommended to prevent chronic disease. However, it has been unclear whether these dietary patterns are equally associated with the health outcomes or not. In a recent study based on three US cohorts (Nurses' Health Study [NHS]/NHS2/Health Professional Follow-up Study [HPFS]), the associations between eight dietary patterns and risk of chronic diseases, defined as a composite of CVD, cancer, and diabetes, were evaluated.[10] In our illustrative example, we aimed to compare the associations of AHEI-2010[5] and rEDIP[6] with the incidence of the chronic disease in the HPFS, which included 43,185 men. These exposures (AHEI-2010, rEDIP) were standardized by its increment from the 10th to the 90th percentile.[10] Person-time of follow-up was calculated from the first available FFQ (19XX) to the occurrence of the chronic disease, death, age 80, or the end of follow-up (January 2016). Cox proportional hazard models were applied to the incidence of chronic disease. Models were stratified by age in months and calendar year of the current questionnaire cycle, and adjusted for physical activity, cigarette smoking status, cigarette smoking pack-years, multivitamin use, regular aspirin use, regular NSAID use, and total energy intake as time-varying covariates, and family history of diabetes, family history of cancer, and family history of CVD as time-fixed covariates.[10] The counting process data structure was used to handle time-varying covariates and left truncation.[11] The time scale for the analysis was then age in months because of the way we structured the data.

First, the associations of each continuous exposure and incident chronic disease were evaluated separately. The hazard ratios (HRs) [95% CI] per increment from the 10th to the 90th percentile were 0.83 [0.79, 0.87] for AHEI-2010 and 0.76 [0.73, 0.80] for rEDIP (**Table 1**). Then, the linear associational patterns between continuous AHEI-2010 and rEDIP were compared (**R code 1**). The coefficient for the interaction term between Exposures and A_type is used for the evaluation. The p-value was 0.005, suggesting that the difference in the associations between the two dietary scores was statistically significant. Although the 95% CIs were overlapped for each exposure, the present test was well powered to detect the difference.

```
#R code 1
```

```r
#First duplicate data with "Exposures" and "A_type" columns. "Exposures" is the exposure value and "A_type" is the exposure type.

data_AHEI = data_original %>% dplyr::mutate(
  Exposures=AHEI_continuous,
  A_type="AHEI"
)

data_rEDIP = data_original %>% dplyr::mutate(
  Exposures=rEDIP_continuous,
  A_type="rEDIP"
)

data_doubled=rbind(data_AHEI,data_rEDIP)

#Then, using the duplicated data, fit a Cox proportional hazard model with an interaction term between "Exposures" and "A_type"

Model = coxph(Surv(Age,Age_at_end_follow_up,outcome) ~ Exposures + covariates + (Exposures + covariates)*A_type + cluster(id) + strata(A_type,Age,period), data=data_doubled)

summary(Model)
```

Second, the exposures were categorized by quintiles. The HRs [95% CI] contrasting the highest (Q5) and least quartile (Q1) were 0.83 [0.78, 0.87] for AHEI-2010, and 0.71 [0.67, 0.75] for rEDIP (**Table 1**). Then, these associational patterns were compared with use of the multivariate Wald test (**R code 2**). The degree of freedom is four, reflecting five categories of the exposures. The p-value was <0.0001, again supportive of the difference in the associational patterns.

```r
#R code 2

#First duplicate data with "Exposures" and "A_type" columns. "Exposures" is the exposure value and "A_type" is the exposure type. Note, the exposure value is factor variable with 5 categories (in R, factor variables are automatically translated into dummy variables in a regression model)

data_AHEI = data_original %>% dplyr::mutate(
  Exposures=AHEI_quintile,
  A_type="AHEI"
)
```

```r
data_rEDIP = data_original %>% dplyr::mutate(
  Exposures=rEDIP_quintile,
  A_type="rEDIP"
)

data_doubled=rbind(data_AHEI,data_rEDIP)

#Then, using the duplicated data, fit two Cox proportional hazard models interaction terms between "Exposures (dummy variables)" and "A_type", and conduct a Multivariate Wald test with degrees of freedom = 4

Model = coxph(Surv(Age,Age_at_end_follow_up,outcome) ~ Exposures + covariates + (Exposures + covariates)*A_type + cluster(id) + strata(A_type,Age,period), data=data_doubled)

#Details of Multivariate Wald test is explained in Figure 3
#A function for multivariate Wald test defined here. The inputs are coxph() fit in the duplicated dataset (ModelFit), Names of categorical exposures (NameofTestExpo), and Name of the interaction term between Exposure and A_type appeared in the summary(Model) (NameofAtype). The function outputs the p-value.

multivariate_wald = function(ModelFit, NameofTestExpo, NameofAtype){
  
  # remove coefficients with "NA" value
  list.NA = which(is.na(coef(ModelFit)))
  coefList.values = coef(ModelFit)[-list.NA]
  coeflist.name = names(coefList.values)
  vcov_coef = vcov(ModelFit)[-list.NA,-list.NA]
  
  # create the test coefficients list
  coeflist.test.name = paste(NameofTestExpo,NameofAtype,sep=':')
  
  # create hypothesis testing comparison matrix
  n_test = length(NameofTestExpo)
  n_coef = length(coeflist.name)
  C_mtx = matrix(rep(0,n_coef*n_test),n_test,n_coef)
  
  for(i in 1:n_test){
    C_mtx[i,] = ifelse(coeflist.name == coeflist.test.name[i], 1, 0)
  }
  
  # multivariate wald test statistics and p-value
  Q = t(C_mtx%*%coefList.values) %*% solve(C_mtx%*%vcov_coef%*%t(C_mtx)) %*% 
    (C_mtx%*%coefList.values)
  p_value = 1-pchisq(Q, n_test)
  
  return(p_value)
}
```

```
NameofTestExpo = c('Exposures2','Exposures3','Exposures4','Exposures5')
NameofAtype = c('A_typerEDIP')
pval = multivariate_wald(Model, NameofTestExpo, NameofAtype)

pval
```

Discussion

In this paper, we described the process of conducting valid statistical tests for the comparison of multiple exposures with respect to a common outcome in the Cox regression models, and we provided sample R codes that implement the proposed approach. In general, existing regression-based separate analysis is unable to account for the potential correlations between exposure coefficients estimated using the same dataset. In contrast, our proposed approach involves a data duplication strategy, where we conduct the traditional analysis on an augmented dataset. Therefore, it provides a convenient way to compare exposure-outcome associations across multiple exposures while taking into account correlations between the estimated associations. Our approach can be extended in other regression-based models, for example including linear, logistic and Poisson regression, where the variance-covariance matrix of the regression coefficient estimates can be estimated using the Sandwich formula accounting for the correlations in the augmented dataset[12], suggestive of the potential for wider applications in epidemiological or medical research. Furthermore, it is not computationally intensive since the calculation do not depend on bootstrapping. In conclusion, the proposed method, based on the duplicated methods, offers a flexible and valid statistical test of multiple exposures for the common outcome with minimal assumptions.

**Footnotes**

# References

1. Henderson SB, Brauer M, Macnab YC, Kennedy SM. Three measures of forest fire smoke exposure and their associations with respiratory and cardiovascular health outcomes in a population-based cohort. Environ Health Perspect. 2011;119:1266–71.

2. Heck WW, Cure WW, Rawlings JO, Zaragoza LJ, Heagle AS, Heggestad HE, et al. Assessing Impacts of Ozone on Agricultural Crops: II. Crop Yield Functions and Alternative Exposure Statistics. Journal of the Air Pollution Control Association. 1984;34:810–7.

3. Sarnat SE, Sarnat JA, Mulholland J, Isakov V, Özkaynak H, Chang HH, et al. Application of alternative spatiotemporal metrics of ambient air pollution exposure in a time-series epidemiological study in Atlanta. J Expo Sci Environ Epidemiol. 2013;23:593–605.

4. Song M, Giovannucci E. Substitution analysis in nutritional epidemiology: proceed with caution. Eur J Epidemiol. 2018;33:137–40.

5. Chiuve SE, Fung TT, Rimm EB, Hu FB, McCullough ML, Wang M, et al. Alternative dietary indices both strongly predict risk of chronic disease. J Nutr. 2012;142:1009–18.

6. Tabung FK, Smith-Warner SA, Chavarro JE, Wu K, Fuchs CS, Hu FB, et al. Development and Validation of an Empirical Dietary Inflammatory Index. J Nutr. 2016;146:1560–70.

7. Korn EL, Graubard BI, Midthune D. Time-to-event analysis of longitudinal follow-up of a survey: choice of the time-scale. Am J Epidemiol. 1997;145:72–80.

8. Lunn M, McNeil D. Applying Cox regression to competing risks. Biometrics. 1995;51:524–32.

9. Wang M, Spiegelman D, Kuchiba A, Lochhead P, Kim S, Chan AT, et al. Statistical methods for studying disease subtype heterogeneity. Stat Med. 2016;35:782–800.

10. Wang P, Song M, Eliassen AH, Wang M, Fung TT, Clinton SK, et al. Optimal dietary patterns for prevention of chronic disease. Nat Med. 2023;29:719–28.

11. Therneau TM. Extending the Cox Model. In: Lin DY, Fleming TR, editors. Proceedings of the First Seattle Symposium in Biostatistics. New York, NY: Springer US; 1997. p. 51–84.

12. Liang K-Y, Zeger SL. Longitudinal data analysis using generalized linear models. Biometrika. 1986;73:13–22.

Table 1. Association of two dietary scores with incident chronic disease in the HPFS and tests for the difference in the associational patterns

|  | Continuous* | Q1 | Q2 | Q3 | Q4 | Q5 |
|---|---|---|---|---|---|---|
| AHEI-2010 | 0.83 [0.79, 0.87] | – | 0.97 [0.92, 1.02] | 0.95 [0.91, 1.01] | 0.91 [0.86, 0.96] | 0.83 [0.78, 0.87] |
| rEDIP | 0.76 [0.73, 0.80] | – | 0.88 [0.84, 0.92] | 0.81 [0.77, 0.86] | 0.77 [0.73, 0.81] | 0.71 [0.67, 0.75] |
| Difference | P = 0.005 | | | P < 0.0001 | | |

Numbers are hazard ratios [95% confidence intervals].

Cox proportional hazard models were fit for N=43,185 participants in the HPFS.. Age was used as the time-scale, and models were stratified by age, calendar year and A_type variable, and adjusted for physical activity, cigarette smoking status, cigarette smoking pack-years, multivitamin use, regular aspirin use, regular NSAID use, total energy intake, family history of diabetes, family history of cancer, and family history of CVD. Statistical tests for the difference in the associational patterns between two exposures were applied for continuous and categorical exposures separately.

*For the continuous exposure analysis, the increment is the difference from the 10th to the 90th percentile of the corresponding diet score

Abbreviations: AHEI, Alternative Healthy Eating Index; rEDIP, reversed empirical dietary inflammatory pattern; HPFS, Health Professional Follow-up Study.

Figure 1

**Original Data**

| ID | A | A' | Y | Time | L1 | L2 | ... |
|---|---|---|---|---|---|---|---|
| 1 | 1 | 1 | 1 | 20 | 1 | | |
| 2 | 0 | 1 | 0 | 19 | 1 | | |
| 3 | 1 | 1 | 0 | 17 | 0 | | |
| 4 | 0 | 0 | 0 | 21 | 0 | | |
| ... | ... | ... | ... | ... | ... | ... | ... |

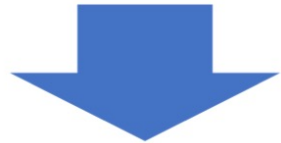

**Original + Duplicated Data**

| ID | Exposures | A_type | Y | Time | L1 | L2 | ... |
|---|---|---|---|---|---|---|---|
| 1 | 1 | A | 1 | 20 | 1 | | |
| 2 | 0 | A | 0 | 19 | 1 | | |
| 3 | 1 | A | 0 | 17 | 0 | | |
| 4 | 0 | A | 0 | 21 | 0 | | |
| ... | ... | ... | ... | ... | ... | ... | ... |
| 1 | 1 | A' | 1 | 20 | 1 | | |
| 2 | 1 | A' | 0 | 19 | 1 | | |
| 3 | 1 | A' | 0 | 17 | 0 | | |
| 4 | 0 | A' | 0 | 21 | 0 | | |
| ... | ... | ... | ... | ... | ... | ... | ... |

***COX model:*** Surv(Time, Y) ~ Exposures + strata(A_type) + L1 + L2 + ... +
Exposures*A_type + L1*A_type + L2*A_type + ... + cluster(ID)

→ Use a coeficient for "Exposures*A_type" for the inference

Figure 2

**Original Data**

| ID | A2 | A3 | A2' | A3' | Y | Time | L1 | L2 | ... |
|---|---|---|---|---|---|---|---|---|---|
| 1 | 1 | 0 | 0 | 1 | 1 | 20 | 1 | | |
| 2 | 0 | 0 | 1 | 0 | 0 | 19 | 1 | | |
| 3 | 0 | 1 | 1 | 0 | 0 | 17 | 0 | | |
| 4 | 0 | 0 | 0 | 0 | 0 | 21 | 0 | | |
| ... | ... | ... | ... | ... | ... | ... | ... | ... | ... |

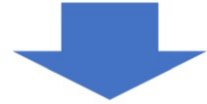

**Original + Duplicated Data**

| ID | Exposures2 | Exposures3 | A_type | Y | Time | L1 | L2 | ... |
|---|---|---|---|---|---|---|---|---|
| 1 | 1 | 0 | A | 1 | 20 | 1 | | |
| 2 | 0 | 0 | A | 0 | 19 | 1 | | |
| 3 | 0 | 1 | A | 0 | 17 | 0 | | |
| 4 | 0 | 0 | A | 0 | 21 | 0 | | |
| ... | ... | ... | ... | ... | ... | ... | ... | ... |
| 1 | 0 | 1 | A' | 1 | 20 | 1 | | |
| 2 | 1 | 0 | A' | 0 | 19 | 1 | | |
| 3 | 1 | 0 | A' | 0 | 17 | 0 | | |
| 4 | 0 | 0 | A' | 0 | 21 | 0 | | |
| ... | ... | ... | A' | ... | ... | ... | ... | ... |

*COX model:* Surv(Time, Y) ~ Exposures2 + Exposures3 + strata(A_type) + L1 + L2 + ... +
Exposures2*A_type + Exposures3*A_type + L1*A_type + L2*A_type + ... + cluster(ID)

→ Multivariate Wald test for 'Exposures2*A_type' and 'Exposures3*A_type'